\documentclass[conference]{IEEEtran}
\IEEEoverridecommandlockouts
\usepackage{cite}
\usepackage{amsmath,amssymb,amsfonts}
\usepackage{algorithmic}
\usepackage{graphicx}
\usepackage{textcomp}
\usepackage{xcolor}
\usepackage{booktabs}
\usepackage{siunitx}
\usepackage{subcaption}
 \usepackage{booktabs}

\usepackage{amsthm}
\usepackage{hyperref}
\usepackage{float}

\DeclareSIUnit{\nauticalmile}{NM}

\newtheorem{theorem}{Theorem} [section]

\theoremstyle{definition}
 
\newtheorem{remark}[theorem]{Remark} 
\newtheorem{example}[theorem]{Example}

\begingroup\expandafter\expandafter\expandafter\endgroup
\expandafter\ifx\csname pdfsuppresswarningpagegroup\endcsname\relax
\else
\fi

\def\BibTeX{{\rm B\kern-.05em{\sc i\kern-.025em b}\kern-.08em
    T\kern-.1667em\lower.7ex\hbox{E}\kern-.125emX}}

\begin{document}

\title{
	QoS based resource management for concurrent operation using MCTS
	\thanks{This project has received funding from the European Union's Preparatory Action on Defence Research under grant agreement No 882407 [CROWN].}
}

\author{
	\IEEEauthorblockN{Sebastian Durst, Kilian Barth, Tobias M\"uller and Pascal Marquardt}
	\IEEEauthorblockA{\textit{Fraunhofer-Institut f\"ur Hochfrequenzphysik und Radartechnik FHR}\\
		Wachtberg, Germany \\
		sebastian.durst@fhr.fraunhofer.de, kilian.barth@fhr.fraunhofer.de
	}
}

\maketitle

\begin{abstract}
	Modern AESA technology enables RF systems to not only perform various radar, communication and electronic warfare tasks on a single aperture, but even to execute multiple tasks concurrently.
	These capabilities increase system complexity and require intelligent or cognitive resource management.
	This paper introduces such a resource management framework based on quality of service based resource allocation and Monte Carlo tree search allowing for optimal system usage and profound decision-making.
	Furthermore, we present experimental verification in a complex application scenario.
\end{abstract}



\section{Introduction}
\label{sec:intro}

Technological advances enable modern and future multifunctional RF-systems (MFRFS) to not only perform radar, communication and electronic warfare (EW) functions on a single aperture, but also to operate concurrently, i.e.\ executing more than one task at a time (cf.~\cite{Heras2022}).
These capabilities require intelligent approaches to resource management to guarantee optimal system performance.

Resource management (i.e.\ task prioritisation, resource allocation and scheduling) is an important part of any modern radar system, as these have to
perform potentially conflicting functions,
and even more so of any cognitive setup.
One important role of the resource management is to select operational parameters from a multitude of possible task configurations differing in resource requirements and resulting utility with the goal to optimise the overall system performance under resource constraints.
A mathematical framework describing this problem is known as \emph{Quality of service based resource allocation model (Q-RAM)}
\cite{Rajkumar1997, Ghosh2006}.
The standard Q-RAM framework, however, is not able to model and efficiently solve the resource allocation problem including concurrent operation, which by the sheer mass of different functions and resulting number of possible combinations is hard to handle.
As MFRFS operate in highly dynamic environments with possibly evolving mission objectives, a pre-selection of useful combinations to reduce complexity seems unsuitable.

In \cite{Marquardt2024} we introduced a suitable resource management framework while concentrating on \emph{modelling} the problem and focused on the creation of appropriate performance models for RF tasks. In the current paper, we specify the actual algorithm and technique \emph{solving} the resource allocation problem with concurrent task execution.
To the best of our knowledge, this is the first publication detailing a flexible solution technique enabling an MFRFS to make use of various concurrent operation modes like multibeam and shared aperture techniques (cf.~\cite{Cox2024}), interleaving or using specialised multi-purpose waveforms.

\section{The Q-RAM problem}
\label{sec:qram_problem}

We will briefly introduce the Q-RAM problem in this section.
The goal of the resource allocation module in a multifunction radar is to maximise the 
\emph{utility} of a set of radar tasks by selecting \emph{operational parameters} (e.g.\ waveform, dwell period or the choice of tracking filter)
while adhering to certain \emph{resource constraints} (e.g.\ radar bandwidth, power) and taking into account the \emph{environmental conditions}, i.e.\ situational data.
A specific choice of operational parameters together with the environmental conditions determine the (expected) \emph{quality} of a task, which is usually task-type related and allows for easier, interpretable user control. Quality and situational data then define the task utility.
For $\{\tau_1,\ldots,\tau_n\}$ a set of radar tasks, $k$ types of resources with resource bounds $R_1,\ldots,R_k$ and environmental conditions $e$, the problem can be formulated as
\begin{align}
	\max_{\phi = (\phi_1,\ldots,\phi_n)}& u(\phi, e)\\
	\textrm{s.t. } \forall j=1,\ldots,k\;\;& \sum_{i=1}^n \big(g_i(\phi_i)\big)_j \leq R_j,
\end{align}
where $\phi_i$ is a configuration for task $\tau_i$, $u$ the system utility and $g_i$ functions mapping task configurations to their resource requirements
(see \cite{Durst2021} for a more detailed description).
The collection of all functions related to a task is called a performance model (cf.~\cite{Marquardt2024}).

\section{A tree structure for concurrent operations}
\label{sec:architecture}

In its standard implementation the Q-RAM algorithm requires an iterative global optimisation process for resource allocation. When dealing with concurrency, this is not suitable any more since the decision whether a task should be executed solely or in combination with other tasks has to be made in advance and is not changeable during the optimisation.
A simple solution would be to only consider fixed (and not all possible) combinations of tasks to be executed concurrently.
Clearly, the result then depends heavily on the exact allowed combinations and thus the problem of choosing beneficial combinations arises naturally.
The goal of the algorithm presented here is to find and evaluate among all possible combinations the most promising ones.
To that end, the space of all possible combinations is given the structure of a tree.
As this tree is in practice to big to be fully searched, an adapted
\emph{Monte Carlo Tree Search} (MCTS) (see~\cite{Browne2012}) approach is pursued.

\begin{figure}[htb] 
	\centering
	\includegraphics[width=0.7\columnwidth]{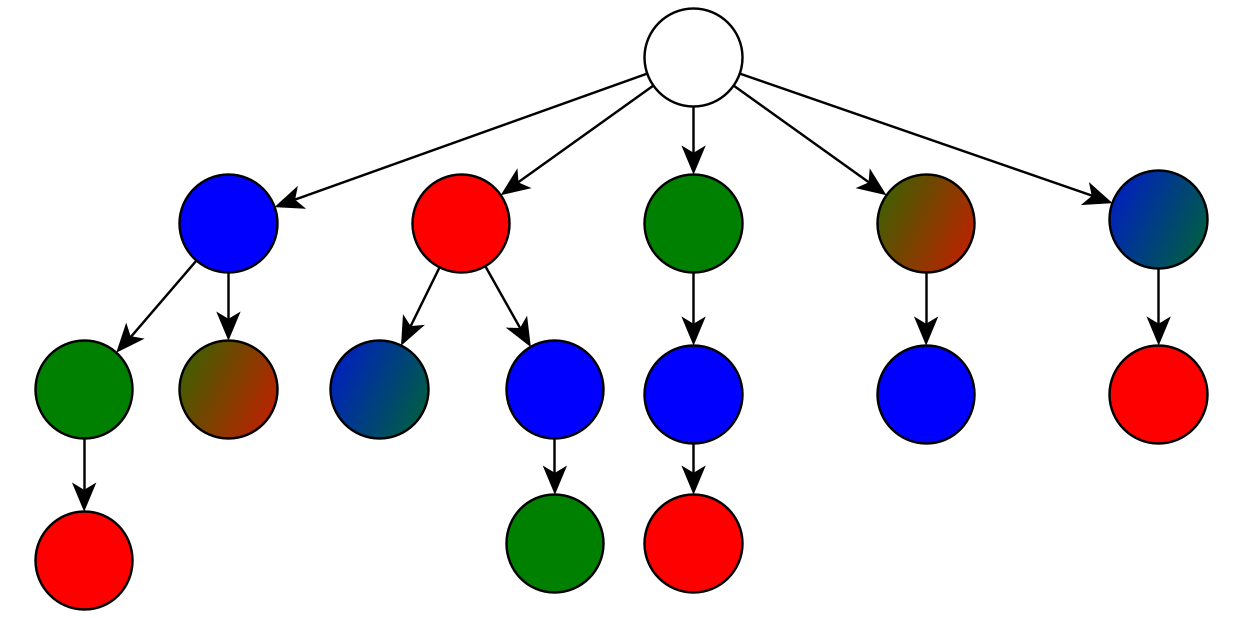}
	\caption{Tree approach. A single colour corresponds to  single mode configuration. Two colours refer to a combination of two tasks.}
	\label{fig:tree}
\end{figure}

The set of all combinations forms a tree in the following way.
Starting from the root, a branch is formed for each task and its combination with other tasks.
This process is done recursively where every task is only allowed to appear once (solely or in combination) inside each path from the root to a leaf
(cf.\ Figure~\ref{fig:tree}).
Note that not necessarily all possible combinations are feasible due to technical restrictions or tactical considerations
like suppressing beams into the same direction working in the same frequency domain,
and the branching can be bounded accordingly.

Starting from a set of tasks $\{T_1,\ldots, T_n\}$, task compatibility can be encoded in a compatibility matrix $C$ with $C_{ij} = 1$ if $i=j$ or $T_i$ can be combined with $T_j$, and $0$ otherwise.
For example, let $T_1$ be a SAR task and $T_2$ a communication task that can be combined, then $C_{12}=1$ indicates that the algorithm is allowed to perform these two tasks concurrently.
Starting from the root, a step down the tree then corresponds to choosing an entry $C_{ij}=1$ and reducing the matrix by setting the $i$-th and $j$-th row and column to zero.
In the example above, this ensures that neither the SAR nor the communication task can be selected again at a later point.
With that notation, leaf nodes are the nodes where the compatibility matrix is zero.
The path to a leaf gives a partition of the set of tasks. This means that as long as suitable performance models for the chosen task combinations are available, a regular Q-RAM optimisation can be performed.
In case the full tree is considered, the leaf with the highest resulting utility is chosen for scheduling.
Note that this method also works if more than two tasks can be combined and executed at the same time by using a higher-dimensional tensor instead of a matrix.

\begin{figure}[htb] 
	\centering
	\includegraphics[width=0.7\columnwidth]{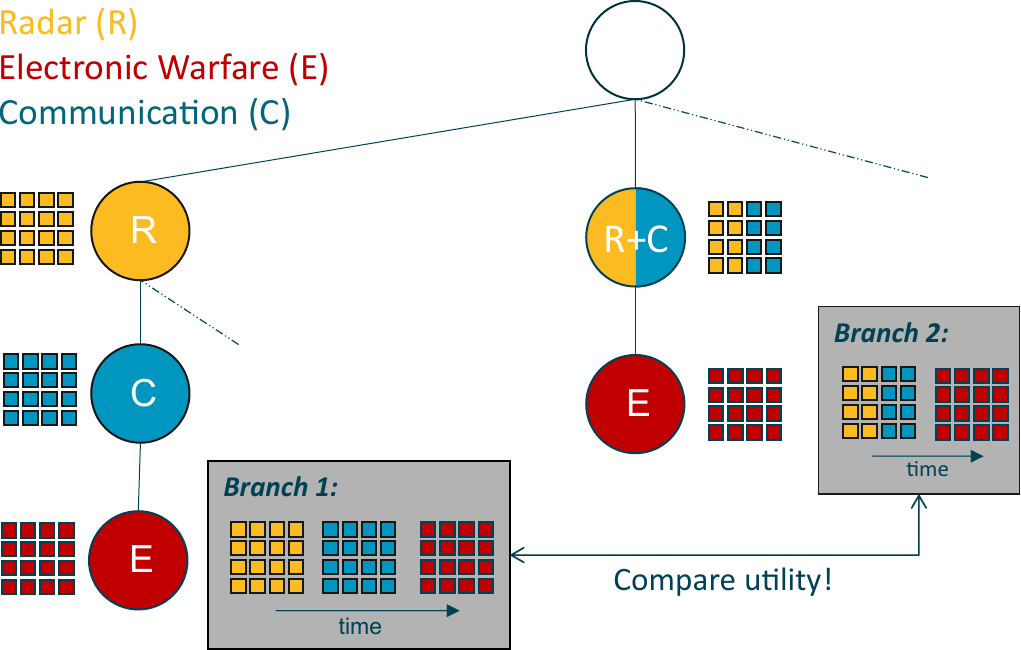}
	\caption{Example of the tree-based optimisation.}
	\label{fig:example}
\end{figure}	

\begin{example}
	Let $T_1=\text{R}$ be a radar, $T_2=\text{E}$ an electronic warfare and $T_3=\text{C}$ a communication task.
	Assume that radar and communication can be performed together in a
	split-aperture mode and no other combinations are allowed.
	Thus, the compatibility matrix is given by
	$$
	C = 
	\begin{pmatrix}
		1 & 0 & 1\\
		0 & 1 & 0\\
		1 & 0 & 1\\
	\end{pmatrix}.
	$$
	The root node then has four children corresponding to the combination of radar and communication $\text{R} + \text{C}$, and any of the individual tasks $\text{R}$, $\text{E}$, $\text{C}$, respectively.
	In the child $\text{R} + \text{C}$, the compatibility matrix is reduced to
	$$
	\widetilde{C} = 
	\begin{pmatrix}
		0 & 0 & 0\\
		0 & 1 & 0\\
		0 & 0 & 0\\
	\end{pmatrix},
	$$
	i.e.\ there is only one possible choice left ($\text{E}$).
	The resulting tree is depicted in Figure~\ref{fig:example}. Each leaf corresponds to a partition of the task set, e.g.\ $\{\text{R} + \text{C}, \text{E}\}$ or $\{\text{R}, \text{E}, \text{C}\}$. For these sets, the Q-RAM problem can be solved and one can choose the overall best option.
\end{example}

In practical applications, the tree is usually too big to be searched completely and a MCTS is used instead.
The key feature of MCTS is to not build the tree completely, but rather to concentrate on the paths that are most promising by exploiting experience but also ensuring sufficient exploration via the Upper Confidence Bound for Trees (UCT) (cf.~\cite{Browne2012}).
In our case, the evaluation uses the resulting Q-RAM utility of a leaf node and in contrast with classical MCTS the best encountered result can be saved within all nodes.
This increases computation speed at the inference stage and further improves results.
Furthermore, by using the regular operation as a baseline,
it is assured that the performance is always at least as good.
One of the biggest advantages is the possibility to stop the search at any time, which is ideal for real-time applications.

\begin{remark}
	There are several different ways to define and understand concurrency in an MFRFS, i.e.\ tasks can be combined and operated concurrently in various manners.
	Among these are dividing the aperture into multiple subapertures (one antenna element is only used in the transmission of a single beam), operating in a true multibeam fashion (a single antenna element can be used in the transmission of multiple beams), MIMO, interleaving in time, and the use of specialised waveforms accomplishing more than one function with a single beam.
	Note that the approach presented here is agnostic to the choice of concurrent operation mode and can even be used to flexibly choose the optimal mode.
\end{remark}

\section{Experimental verification}
\label{sec:sim}

The presented framework has been validated by comparing its performance with standard non-concurrent operation (using a state-of-the-art Q-RAM based resource manager) in a relevant scenario
via a Monte Carlo simulation using Fraunhofer FHR's Cognitive Radar Simulator (CoRaSi).

\subsection{Scenario description}

\begin{figure}
	\centering
	\includegraphics[width=0.85\columnwidth]{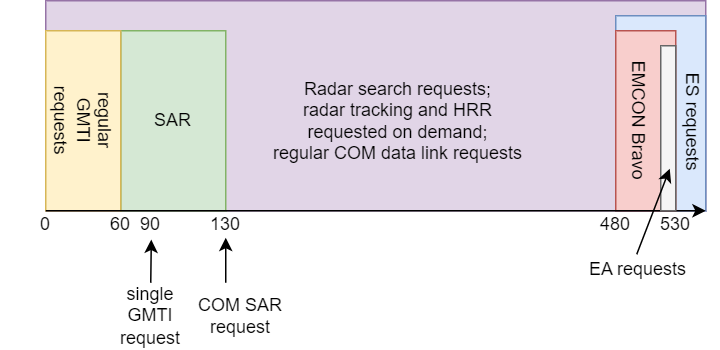}
	\caption{Graphical representation of the storyboard.}\label{fig:storyboard}
\end{figure}

The presented scenario has been derived from the CONOPS of the CROWN project.
The simulated RF system is mounted on an unmanned combat aerial vehicle (UCAV) which is part of a mission (of multiple platforms) to identify and destroy a ballistic missile launcher located \SI{300}{\nauticalmile} behind the forward line of allied troops.
Its exact position is unknown beforehand and has to be identified via the use of synthetic aperture radar (SAR) imagery. The resulting image has to be sent to a ground station for further analysis.
Both the SAR and the communication task are essential for mission success.
The UCAV has to perform  a total of ten different RF modes for gathering information and ensure self-protection in enemy airspace.
The simulated phase of the scenario has a duration of \SI{550}{\s}. The general outline describing a timeline of task requests is given in Figure~\ref{fig:storyboard}.
In the depicted storyboard there are some tasks that are requested on a regular basis like surveillance, tracking or the communication data link.
Other tasks like GMTI or SAR are requested at specific times.
In standard operation mode, the requested SAR task (starting at around \SI{60}{\s}) has a long duration in which it blocks the antenna for other tasks.
At around \SI{480}{\s} emission control (EMCON) level Bravo is enforced, i.e.\ RF transmissions are prohibited.
Note that a task request does not necessarily mean that the task is going to be executed but only that there is a demand for it.
Whether a task is executed is decided upon the resource manager, which also takes into account other requests, environmental conditions and the mission profile.
For a more detailed description of the scenario the reader is referred to~\cite{Marquardt2024}.

\subsection{FHR Cognitive Radar Simulator CoRaSi}

\begin{figure}[t] 
	\centering
	\includegraphics[width=.39\textwidth]{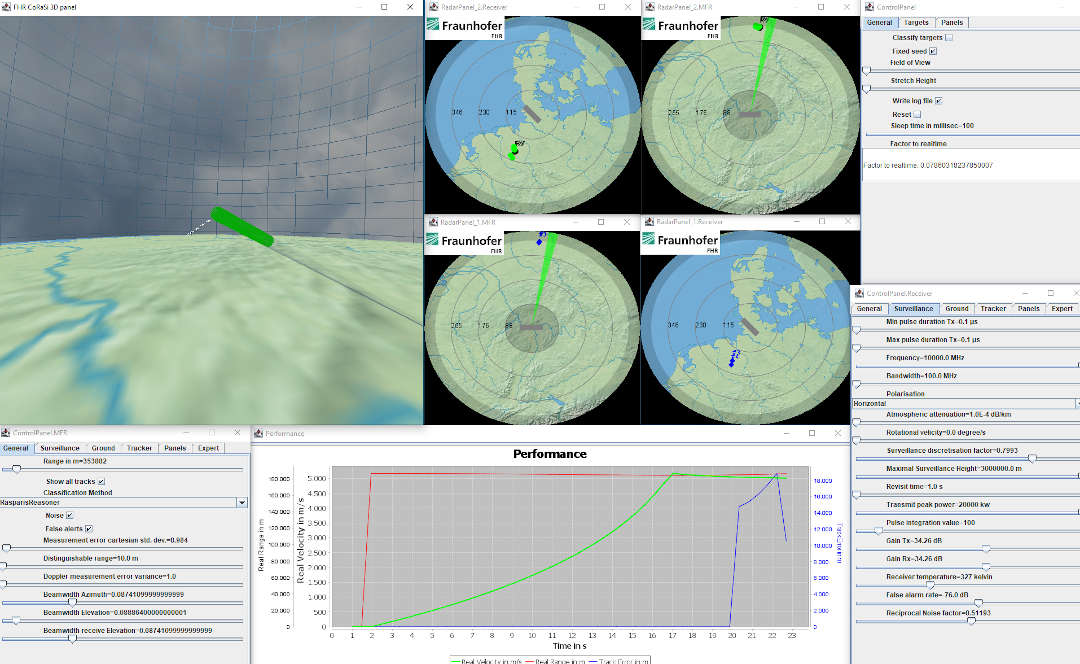}
	\caption{Graphical user interface of CoRaSi.} \label{fig:corasi}
\end{figure}

Fraunhofer FHR's Cognitive Radar Simulator (CoRaSi) (cf.\ Figure~\ref{fig:corasi}) is a multi-purpose software simulator for phased array radars with electronic beam steering written in Java, which was developed as part of the basic funding by the German Ministry of Defence.
It enables real-time analyses of radar systems in different frequency ranges, rotating or static, with arbitrary antenna patterns and search strategies. The simulation includes the most important functions such as search, tracking, target classification, resource management and data fusion across multiple sensors.
CoRaSi was originally developed for ground-based multifunction radars equipped with AESA antennas,
but is now also capable of simulating ship and air-based systems.
In case more than one sensor is simulated, different fusion concepts (central, decentral, hybrid) are available to merge information.
Due to its variability CoRaSi is able to analyse many phased array radars based on given parameters.
For more details, see \cite{Mueller2023}.

\subsection{Experimental results}

The MCTS-based resource management framework was tested for a split-aperture multioperation mode, in which two tasks can be executed concurrently by dividing the aperture, and compared with standard non-concurrent operation. All results are based on \num{25} randomised Monte Carlo simulations each.
Although the primary mission (SAR imaging to identify the launcher) is accomplished in both cases, the concurrent operation clearly outperforms standard operation, as can be seen by the significantly higher total utility, i.e.\ system performance (see Table~\ref{tab:util}).
This means that with concurrent operation, subordinate or secondary mission goals -- e.g.\ reconnaissance of the scenery -- can be achieved more successfully.
One main driver for this is the better tracking performance, which deviates between the two modes under consideration in particular between \SI{60}{\s} and \SI{130}{\s} into the simulation when the system has to perform the stripmap SAR task.
Although there is a slight increase of roughly \SI{100}{\m} in track error also for concurrent operation, the overall track error is kept well under control in this mode. On the contrary, in standard operation the mean track error increases to more than \SI{3000}{\m} with the third quantile being as high as \SI{6000}{\m}.

\begin{table}
	\centering
	\caption{Comparison of total system utilities. Results are averaged over 25 simulation runs.}\label{tab:util}
	\begin{tabular}{lrrrrr}
			\toprule
			{} &  median &   min &     max &    mean &  std. dev. \\
			\midrule
			standard &  860.66 & 858.16 & 864.25 &  860.96 &     1.59 \\
			multioperation   &  879.07 & 875.74 & 884.32 & 879.61 &     2.15 \\
			\bottomrule
		\end{tabular}
\end{table}

This is not surprising, as the standard mode does not have the ability to actively track targets during a SAR execution, while in multioperation mode SAR imaging and tracking can be combined.
As the track error increases over time if the target is not illuminated, the benefit of concurrent operation becomes immediately apparent.
For other task types the positive effects of concurrency might be less visible at first sight.
For example, a requested high resolution range profile (HRRP) classification task will usually not be deleted when the aperture is blocked by
another more important task, but rather postponed to a later execution time.
Nevertheless, the system can benefit from early execution as potential threats can be identified sooner, leaving more time for counter-measures.
The same is true for communication and electronic surveillance as well as regular radar search.
The effects of early warning and threat detection are often not easy to measure except in extreme cases but an enhanced situational awareness is essential for successful mission execution.

\begin{figure}
	\centering
	\includegraphics[width=0.85\columnwidth]{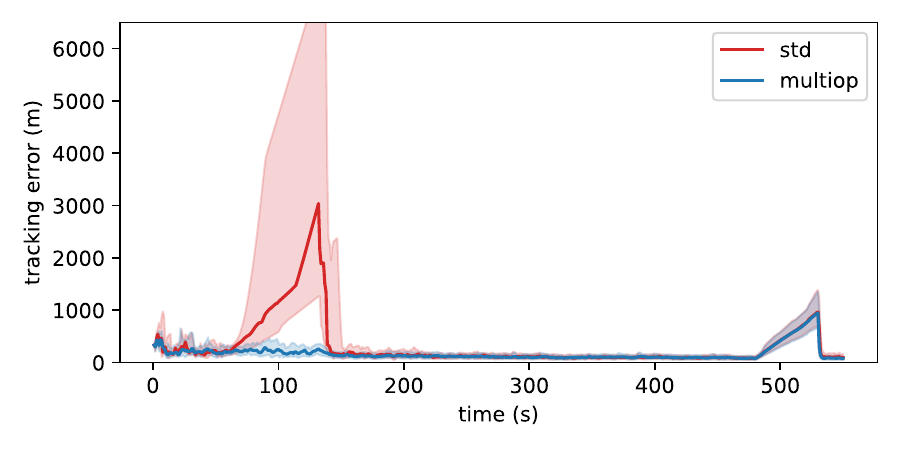}
	\caption{Track error over time.}\label{fig:track_error}
\end{figure}

\begin{figure}
	\centering
	\includegraphics[width=0.7\columnwidth]{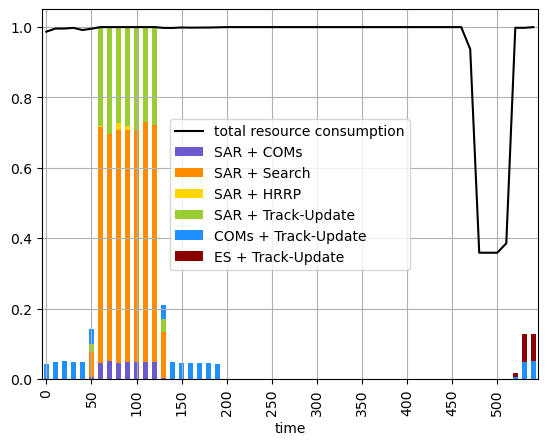}
	\caption{Share of concurrent operation by task type over time in multioperation mode.}\label{fig:used_concurrency}
\end{figure}

The actual use of the concurrency is depicted in Figure~\ref{fig:used_concurrency}. As can be seen, the resource manager selects various task combinations for concurrent execution -- this is decided upon by the MCTS based resource allocation algorithm depending on the current situation with the goal to maximise system performance.
The system does not only make use of its concurrent capabilities during phases of high load (as during SAR execution) but also in the periods immediately afterwards, e.g.\
between 130s and 200s and after 530s, which are the timespans after the resource-intensive SAR imaging and after the EMCON restriction, respectively, as the number of tasks to fulfil as well as their urgency increases.
This enables the system to quickly get all functions which might have suffered during the heavy load phase back to normal performance levels.

In summary, the quality-of-service based resource management using MCTS enables an efficient concurrent operation which does not only improve utility under heavy load but is also able to equalise the after-effects more quickly. Although not being used at times of low loads, the performance of the concurrent modes is considerably better overall in the simulated scenario, which poses a typical mixture of times of high and low demand.

\section{Conclusion}\label{sec:conclusion}

An enhancement of a Q-RAM resource management has been presented extending its applicability to general multifunction RF-systems with concurrent task execution.
The framework is based on an MCTS approach and is agnostic to the technical implementation of concurrency, including split aperture operation.
A resource management with the described capabilities is a prerequisite for future and cognitive MFRFS and the given experimental results showcase its benefit over traditional operation modes in challenging and realistic scenarios.

\section*{Acknowledgment}
The authors would like to thank the other partners of the PADR project "European active electronically scanned array with Combined Radar, cOmmunications, and electronic Warfare fuNctions for military applications" (CROWN) for the fruitful collaboration.


\bibliographystyle{IEEEtran}
\bibliography{IEEEabrv,lit}

\begin{thebibliography}{1}
\providecommand{\url}[1]{#1}
\csname url@samestyle\endcsname
\providecommand{\newblock}{\relax}
\providecommand{\bibinfo}[2]{#2}
\providecommand{\BIBentrySTDinterwordspacing}{\spaceskip=0pt\relax}
\providecommand{\BIBentryALTinterwordstretchfactor}{4}
\providecommand{\BIBentryALTinterwordspacing}{\spaceskip=\fontdimen2\font plus
\BIBentryALTinterwordstretchfactor\fontdimen3\font minus
  \fontdimen4\font\relax}
\providecommand{\BIBforeignlanguage}[2]{{%
\expandafter\ifx\csname l@#1\endcsname\relax
\typeout{** WARNING: IEEEtran.bst: No hyphenation pattern has been}%
\typeout{** loaded for the language `#1'. Using the pattern for}%
\typeout{** the default language instead.}%
\else
\language=\csname l@#1\endcsname
\fi
#2}}
\providecommand{\BIBdecl}{\relax}
\BIBdecl

\bibitem{Heras2022}
M.~Heras \emph{et~al.}, ``Crown project, towards a european multifunction aesa
  system,'' in \emph{2022 IEEE International Symposium on Phased Array Systems
  \& Technology (PAST)}.\hskip 1em plus 0.5em minus 0.4em\relax IEEE, 2022, pp.
  1--8.

\bibitem{Rajkumar1997}
R.~Rajkumar, C.~Lee, J.~Lehoczky, and D.~Siewiorek, ``A resource allocation
  model for qos management,'' in \emph{Proceedings Real-Time Systems
  Symposium}.\hskip 1em plus 0.5em minus 0.4em\relax IEEE, 1997, pp. 298--307.

\bibitem{Ghosh2006}
S.~Ghosh, R.~Rajkumar, J.~Hansen, and J.~Lehoczky, ``Integrated qos-aware
  resource management and scheduling with multi-resource constraints,''
  \emph{Real-Time Systems}, vol.~33, pp. 7--46, 2006.

\bibitem{Marquardt2024}
P.~Marquardt, S.~Durst, K.~Barth, and T.~Müller, ``A resource management
  approach for concurrent operation of rf functionalities,'' in \emph{2024
  International Radar Conference (RADAR2024)}.\hskip 1em plus 0.5em minus
  0.4em\relax IEEE, 2024, pp. 1--6.

\bibitem{Cox2024}
P.~Cox and W.~van Rossum, ``Radar resource management for active tracking using
  split-aperture phased arrays,'' in \emph{2024 IEEE Radar Conference
  (RadarConf24)}, 2024, pp. 1--6.

\bibitem{Durst2021}
S.~Durst and S.~Br{\"u}ggenwirth, ``Quality of service based radar resource
  management using deep reinforcement learning,'' in \emph{2021 IEEE Radar
  Conference (RadarConf21)}.\hskip 1em plus 0.5em minus 0.4em\relax IEEE, 2021,
  pp. 1--6.

\bibitem{Browne2012}
C.~B. Browne \emph{et~al.}, ``A survey of monte carlo tree search methods,''
  \emph{IEEE Transactions on Computational Intelligence and AI in games},
  vol.~4, no.~1, pp. 1--43, 2012.

\bibitem{Mueller2023}
T.~Müller, S.~Durst, P.~Marquardt, and S.~Brüggenwirth, ``Quality of service
  based radar resource management for navigation and positioning,'' in
  \emph{2023 IEEE/ION Position, Location and Navigation Symposium (PLANS)},
  2023, pp. 59--66.

\end{thebibliography}

\end{document}